# Fully 3D Implementation of the End-to-end Deep Image Prior-based PET Image Reconstruction Using Block Iterative Algorithm

Fumio Hashimoto, Yuya Onishi, Kibo Ote, Hideaki Tashima, and Taiga Yamaya

*Abstract*—Deep image prior (DIP) has recently attracted attention owing to its unsupervised positron emission tomography (PET) image reconstruction, which does not require any prior training dataset. In this paper, we present the first attempt to implement an end-to-end DIP-based fully 3D PET image reconstruction method that incorporates a forward-projection model into a loss function. To implement a practical fully 3D PET image reconstruction, which could not be performed due to a graphics processing unit memory limitation, we modify the DIP optimization to block-iteration and sequentially learn an ordered sequence of block sinograms. Furthermore, the relative difference penalty (RDP) term was added to the loss function to enhance the quantitative PET image accuracy. We evaluated our proposed method using Monte Carlo simulation with [$^{18}$F]FDG PET data of a human brain and a preclinical study on monkey brain [$^{18}$F]FDG PET data. The proposed method was compared with the maximum-likelihood expectation maximization (EM), maximum-a-posterior EM with RDP, and hybrid DIP-based PET reconstruction methods. The simulation results showed that the proposed method improved the PET image quality by reducing statistical noise and preserved a contrast of brain structures and inserted tumor compared with other algorithms. In the preclinical experiment, finer structures and better contrast recovery were obtained by the proposed method. This indicated that the proposed method can produce high-quality images without a prior training dataset. Thus, the proposed method is a key enabling technology for the straightforward and practical implementation of end-to-end DIP-based fully 3D PET image reconstruction.

## I. Introduction

POSITRON emission tomography (PET) is a powerful noninvasive imaging modality that can measure glucose metabolism and other functions using various PET tracers in a living body [1]. However, measured PET data are often affected by statistical noise due to the limitations of the PET tracer dose and acquisition time, which leads to poor performance in small lesion detection and low quantitative accuracy of pharmacokinetic properties. To date, a variety of methods, such as post-processing denoising [2]–[5] and image reconstruction [6]–[8], have been proposed to overcome these limitations.

In PET image reconstruction, various regularizations have been used to suppress statistical noise and improve PET image quality. A common strategy for penalized PET image reconstruction is to measure the spatial smoothness over the PET image space through Gibbs priors [9]–[12]. Alternatively, several anatomical-guided PET image reconstructions, using computed tomography (CT) and magnetic resonance (MR) images, which penalize the prior anatomical information space have been proposed [13]–[15]. In addition to these conventional algorithms, some deep learning-based PET image reconstructions have recently been proposed to achieve reasonable performance [16]–[19].

The first attempt at deep learning-based medical image reconstruction was the automated transform by manifold approximation (AUTOMAP) by Zhu et al. in 2018, which uses a deep neural network to learn the direct mapping from the measured data space to the reconstructed imaging space [20]. The advantage of this method is that high-quality reconstructed images can be quickly obtained. Starting with the AUTOMAP method, several convolutional neural network (CNN)-based direct image reconstruction methods have been proposed for PET imaging [21]–[25]. Another approach to deep learning-based PET image reconstruction is regularization reconstruction, which uses a CNN representation as prior information to push the limit of existing PET image reconstruction methods [26,27]. Gong et al. proposed iterative PET image reconstruction using CNN representation as a regularizer, which can solve the expectation maximization (EM) reconstruction and CNN denoising separately [26]. Although these data-driven approaches exhibit excellent reconstructed image characteristics, network training still requires large datasets and depends on the quality of the training dataset, which may cause an assortment of clinical issues, such as generalization performance in different PET tracers and diseases that are not included in the training dataset.

To overcome these drawbacks, Ulyanov et al. proposed a deep image prior (DIP) that utilizes the CNN structure itself as an intrinsic regularizer [28]. Following the success of natural image enhancements using DIP, several PET image denoising and reconstruction methods using the DIP framework have been proposed [29]-[33]. Gong et al. [34] and Ote et al. [35] further explored deep learning-based iterative PET image reconstruction using DIP regularization by utilizing the same strategy as in [26]. Furthermore, we previously developed a simple implementation of an end-to-end PET image reconstruction algorithm using DIP and a forward projection model. The algorithm was formulated as a single optimization problem for neural network training and can be implemented with fewer hyperparameters than the other deep learning-based iterative PET image reconstruction [36]. However, it is restricted to 2D PET data because the tensor gradients of all forward projection processes must be stored in a graphics processing unit (GPU) memory.

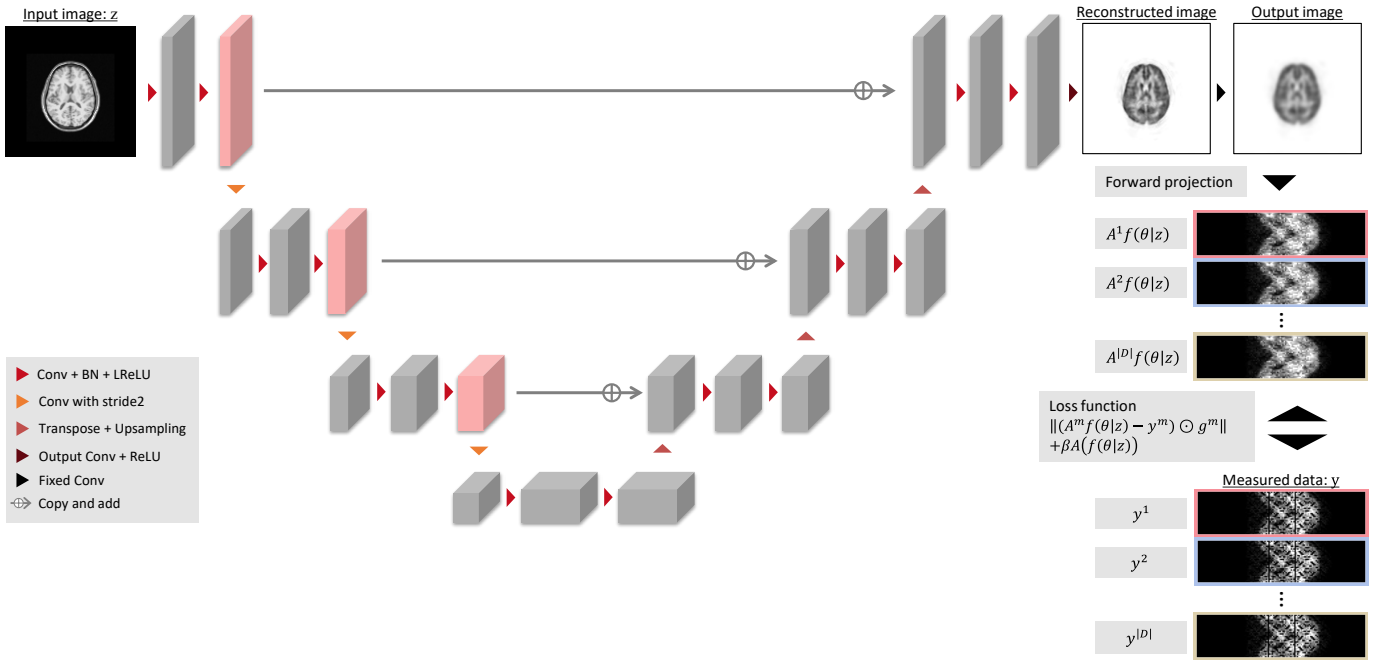

Fig. 1. Implementation overview of the proposed end-to-end DIP-based PET image reconstruction. The proposed method was performed in the following steps; (1) The conditional vector (MR image) z was input into the network. (2) The reconstructed PET image was obtained from the network output. (3) The block sinogram was calculated from the output image through a fixed convolution layer that simply incorporates the blurring model of the system. (4) The L2 loss was calculated with each measured block sinogram and estimated block sinogram, and the network was learned sequentially by mini-batch optimization.

The contributions of this paper is that we 1) propose a novel end-to-end DIP-based fully 3D PET image reconstruction that incorporates a forward projection model into a loss function; 2) modify the DIP optimization to block-iteration and sequentially learn an ordered sequence of block sinograms to implement a practical fully 3D PET image reconstruction; and 3) add a further penalty term, relative difference penalty (RDP) [12], in the loss function to enhance the quantitative accuracy of the reconstructed PET image without compromising the visual quality of the PET image. To the best of our knowledge, this is a novel approach in the application of deep learning-based end-to-end image reconstruction to fully 3D PET data. We quantitatively compared the results of the proposed method with those of other iterative PET image reconstruction algorithms using Monte Carlo simulation and preclinical data from the monkey brain [$^{18}$F]FDG PET scan.

## II. METHODOLOGY

### A. Conditional deep image prior

The DIP framework proposed by Ulyanov et al. [28] is an unsupervised method that utilizes a CNN structure itself as an intrinsic regularizer. Therefore, DIP can free the user from the preparation of large training datasets.

The learning of the DIP denoising task only uses a single pair of random-noise input data, $z$, and a target image, $x_0$ degraded by statistical noise - such as Gaussian and Poisson noise.

$$\theta^* = \underset{\theta}{\operatorname{argmin}} E(x_0; f(\theta|z)),$$
$$x^* = f(\theta^*|z), \qquad (1)$$

where $f$ represents a CNN structure with trainable parameters $\theta$. $E(\cdot)$ is an objective function, such as the mean squared error (MSE) or mean absolute error. The DIP denoising task should stop the learning process with moderate iterations before overfitting the noise components in the target image.

For the PET image denoising task, in addition to the intrinsic prior of the CNN structure, the patient's anatomical information, such as MR and CT images, can be used as an additional condition to enhance the PET image quality. Here, conditional DIP can be performed by replacing $z$ in (1) with a conditional vector of anatomical information.

### B. PET imaging model

In general, the PET imaging model can express that the projection data $y \in \mathbb{R}^{M \times 1}$ is related to the spatial distribution of the PET tracer $x \in \mathbb{R}^{N \times 1}$ through the Radon transformation

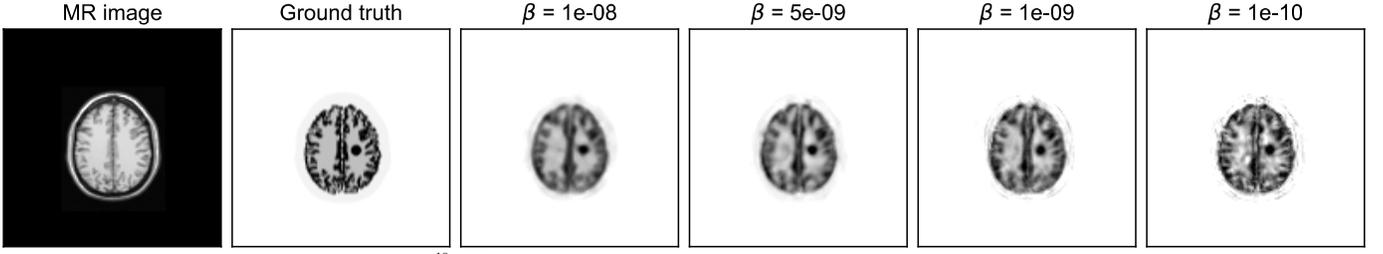

Fig. 2. Reconstruction results of the human brain [$^{18}$F]FDG simulation data for different regularization parameter $\beta$ with subsets 2. The magnified images of the red squared region are shown in the bottom row. Each PSNR corresponding to these algorithms is listed in the titles of each algorithm.

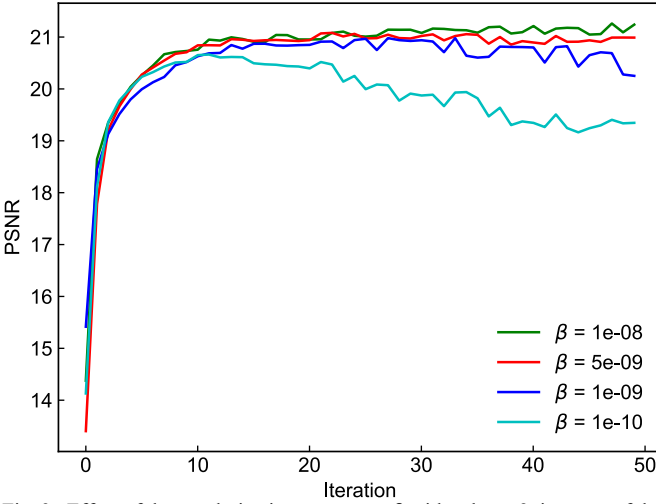

Fig. 3. Effect of the regularization parameter $\beta$ with subsets 2, in terms of the PSNR for the proposed DIP-based reconstruction method.

using the system matrix $A \in \mathbb{R}^{M \times N}$,

$$y = Ax. \qquad (2)$$

The system matrix $A$ projects from the PET image $x$ to the projection data $y$. $M$ and $N$ are the number of lines of response and voxels in the image space, respectively.

### C. End-to-end DIP-based 3D PET image reconstruction

In the proposed end-to-end DIP-based PET image reconstruction, the reconstructed PET image $x$ is defined as the output of the conditional DIP representation as follows:

$$x = f(\theta|z). \qquad (3)$$

The PET image $x$ can be obtained by solving the constrained optimization as follow,

$$\min E(y, Ax) \\ \text{s.t. } x = f(\theta|z). \qquad (4)$$

To obtain the PET image $x$ in an end-to-end manner, we substitute the constraint term into the objective function and solve this for the trainable parameters $\theta$.

$$\theta^* = \underset{\theta}{\mathrm{argmin}}\, E(y, Af(\theta|z)), \\ x^* = f(\theta^*|z). \qquad (5)$$

To enhance the quantitative accuracy of the reconstructed PET image and prevent overfitting in the DIP framework, we introduce the RDP term into the loss function as a penalization for the DIP optimization.

$$\theta^* = \underset{\theta}{\mathrm{argmin}}\, E(y, Af(\theta|z)) + \beta R(f(\theta|z)), \qquad (6)$$

$$R(x) = \sum_j \sum_{k \in N_j} \frac{(x_j - x_k)^2}{(x_j + x_k) + \gamma|x_j - x_k|}, \qquad (7)$$

where $R$ is the RDP that penalizes the differences between neighboring voxels in the PET image domain, and $\beta$ and $\gamma$ determine the regularization parameter and shape of the function in the RDP, respectively. The RDP performs a higher degree of smoothing in relatively low-activity areas and a lower degree in high-activity areas in the PET image space. The RDP has been reported to improve contrast recovery and reduce background noise in phantom as well as clinical evaluations, compared to the OSEM algorithm [37]. In this study, we used $\gamma = 2$ - a condition used for typical clinical PET scanners.

In the 3D PET data, this optimization problem cannot be implemented on current GPU boards because of the large amount of data – including, not only 3D projection data, conditional vector, and the network parameters, but also calculation processes of the 3D forward projection to keep gradients to optimize the trainable parameters through backward propagation. To reduce GPU memory usage, the optimization in (6) can be rewritten as block-iteration and can sequentially learn an ordered sequence of block sinograms as follows:

$$\theta^* = \underset{\theta}{\mathrm{argmin}} \sum_{d \in D} E(y^d, A^d f(\theta|z)) + \beta R(f(\theta|z)), \qquad (8)$$

where $D$ is the ordered subsets of the 3D sinogram, and $d$ is the access order of the subsets. Additionally, we store system matrix $A$ using a sparse matrix in a coordinate list format to compress its empty part.

In this study, MSE is used for the objective function, and we apply a binary mask to ignore the loss calculation of the detector gap in the sinogram space. The final equation of the optimization function (8) is expressed as follows:

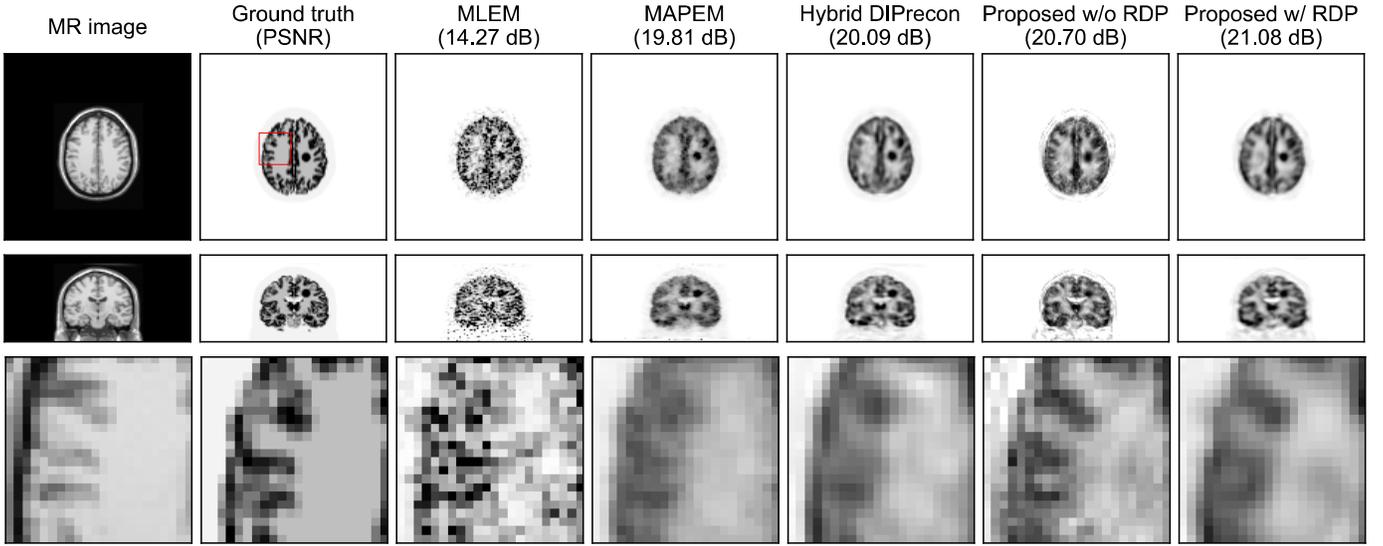

Fig. 4. Reconstruction results of the human brain [$^{18}$F]FDG simulation data for different reconstruction algorithms; the MLEM, MAPEM [12], hybrid DIPrecon [34], proposed method without RDP (Proposed w/o RDP), and proposed method with RDP (Proposed w/ RDP) (left-to-right). The magnified images of the red squared region are shown in the bottom row. Each PSNR corresponding to these algorithms is listed in the titles of each algorithm.

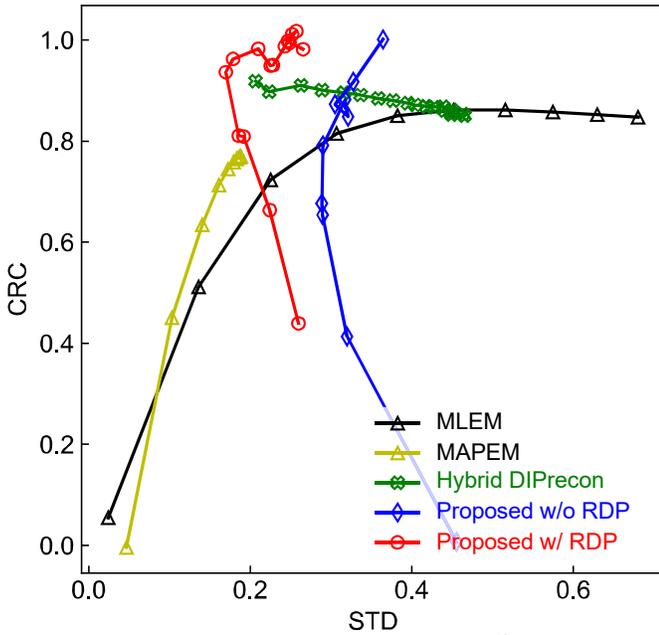

Fig. 5. CRC-STD tradeoff curves of the human brain [$^{18}$F]FDG simulation data. Markers are plotted every ten iterations from one to 100 in the MLEM and MAPEM, every ten iterations from one to 200 in the hybrid DIPrecon (left to right), every two epochs from 1 to 20 in the proposed method without RDP, and every three epochs from 2 to 50 in the proposed method with RDP (lower left to upper right).

$$\theta^* = \underset{\theta}{\arg\min} \sum_{d \in D} \left\| (A^d f(\theta|z) - y_0^d) \odot g^d \right\|^2 + \beta R(f(\theta|z))$$
$$x^* = f(\theta^*|z), \qquad (9)$$

where $g \in \{0,1\}^{M \times 1}$ represents the binary mask corresponding to the detector gap between each detector module, and $\odot$ is the Hadamard product, which works as an inpainting task of the detector gap in the sinogram space. In this study, an MR image was used as the conditional vector $z$.

An implementation overview of the proposed end-to-end PET image reconstruction method is shown in Fig. 1. To summarize, the proposed method is performed in the following steps. (1) The conditional vector (MR image) $z$ is input into the network. (2) The reconstructed PET image is obtained from the network output. (3) The block sinogram is calculated from the output image through a fixed convolution layer that simply incorporates the blurring model of the system. (4) The L2 loss is calculated with each measured block sinogram and estimated block sinogram, and the network is learned sequentially by mini-batch optimization.

### D. Network Architecture

A typical 3D U-Net-based structure was used in this study [38]. The network consists of three parts: 1) encoding, 2) decoding, and 3) system modeling. 1) The encoding part consists of three CNN segments that repeated the following components: two $3 \times 3 \times 3$ convolution layers with batch normalization (BN), a leaky rectified linear unit (LReLU) activation, and a $3 \times 3 \times 3$ downsampling convolution layer. The number of feature maps doubled and the number of rows, columns, and slices decreased by half for each downsampling layer. 2) The decoding part also consists of three CNN segments that repeat the following components: a $3 \times 3 \times 3$ transpose convolution layer with trilinear upsampling and concatenation by adding the corresponding feature maps from the encoding part and two $3 \times 3 \times 3$ convolution layers with BN and LReLU activation. Then, a $1 \times 1 \times 1$ convolution layer with a rectified linear unit (ReLU) activation was performed to output the reconstructed PET image. 3) The point spread function is represented as a fixed 3D convolution layer after the network output to incorporate the blurring effect of the system model and uses $\sigma = 0.5$ voxel Gaussian kernel in this study.

Some DIP-based PET imaging applications of PET denoising and image reconstruction introduced the limited-memory Broyden–Fletcher–Goldfarb–Shanno (L-BFGS) algorithm [39] as an optimizer. The L-BFGS algorithm is a quasi-Newton method that uses the second-order gradient and converges faster than other first-order optimizers, such as the stochastic gradient descent and Adam algorithms [30],[33]-[35]. In this study, we optimized (9) using the Stochastic L-BFGS algorithm extended to a mini-batch optimization [40].

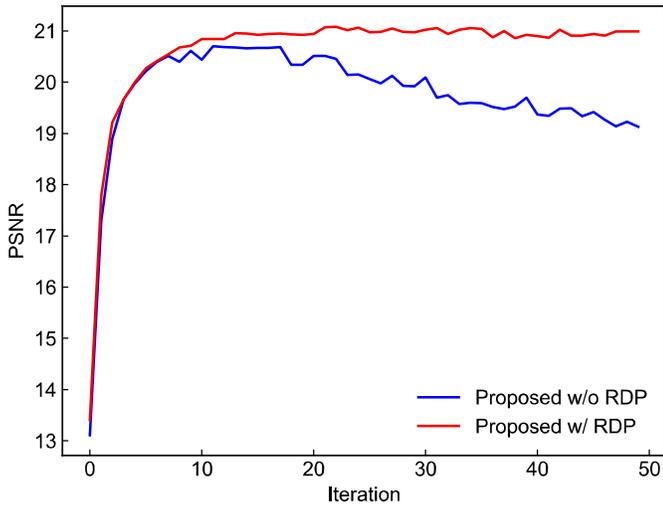

Fig. 6. Impact of the RDP in terms of the PSNR for the proposed DIP-based reconstruction method.

In this study, we used a workstation with PyTorch 1.12.1 (https://pytorch.org/) and a GPU board of NVIDIA A100 with 80 GB memory.

## III. EXPERIMENTAL SETUP

We evaluated the proposed method with other iterative PET image reconstruction algorithms using Monte Carlo simulation of a human brain phantom with a contrast corresponding to [$^{18}$F]FDG and preclinical PET data of a brain with an [$^{18}$F]FDG injection.

### A. Simulation data

The numerical phantom was used from BrainWeb (https://brainweb.bic.mni.mcgill.ca/brainweb/), and the projection data was generated using our own Monte Carlo simulation code with the geometry of the brain-dedicated PET scanner [41]. The diameter of the detector ring was 486.83 mm, consisting of 28 and 4 detector units in the ring and axial directions, respectively. Each detector unit consisted of 16 × 16 cerium-doped lutetium–yttrium oxyorthosilicate crystals. The crystal size was 3.14 × 3.14 × 20 mm$^3$. The radioactive contrast of the gray matter, white matter, and cerebrospinal fluid (CSF) was set to a typical [$^{18}$F]FDG contrast (1:0.25:0.05). The attenuation coefficients were set to 0.00958 mm$^{-1}$ in the soft tissue and CSF, and 0.0151 mm$^{-1}$ in the bone. The simulation included attenuation and scatter effects without positron range, angular deviation, and random events. Scatter events were removed from the list data to simplify the simulation. The simulated 3D PET sinogram was generated from list-mode data using the nearest neighbor interpolation; the ring difference was binned with a span of seven. The sizes of the sinogram and reconstructed image were 128 (bin) × 128 (angle) × 64 (slice) × 19 (oblique angle) and 128 × 128 × 64 voxels with a voxel size of 3.0 × 3.0 × 3.2 mm$^3$, respectively. The projection data had 2,921,540 counts. As a preprocessing step, we applied component-based normalization and attenuation correction to the simulated sinogram.

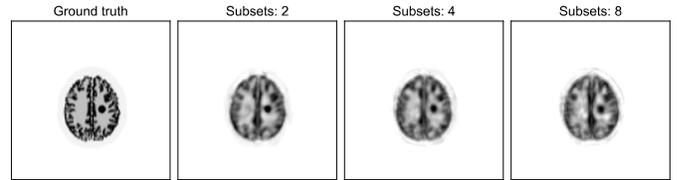

Fig. 7. Reconstruction results of the human brain [$^{18}$F]FDG simulation data for different number of subsets.

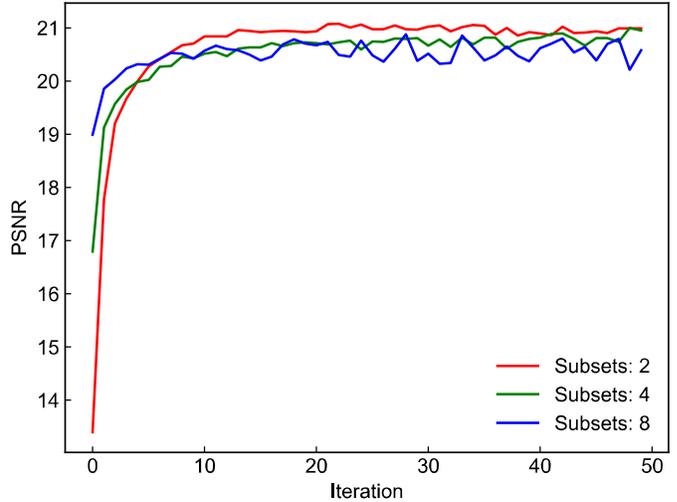

Fig. 8. Effect of the number of subsets in terms of the PSNR for the proposed DIP-based reconstruction method.

### B. Preclinical data

This preclinical study was approved by the Animal Ethics Committee of the Central Research Laboratory, Hamamatsu Photonics K.K. We scanned the brain of a conscious rhesus monkey using an animal PET scanner (SHR-38000, Hamamatsu Photonics K.K., Japan). Before the emission scan, a 30-minute transmission scan was performed using a $^{68}$Ge–$^{68}$Ga rod source. A 30-minute emission scan was started 60 min after bolus injection of [$^{18}$F]FDG with 194.7 MBq. For preprocessing, scatter and attenuation corrections were performed using the convolution subtraction method [42] and the reprojection of the transmission image into the measured sinogram space. The low-count emission data were generated by periodically downsampling to 1/20th of the acquired list-mode data. The sizes of the sinogram and reconstructed image were the same as the simulation data, except for the reconstructed voxel size of 0.65 × 0.65 × 1.0167 mm$^3$.

The conditional vector of the T1-weighted MR image was scanned on another day and manually registered to the emission image by two radiological technologists.

### C. Evaluation

We compared the proposed method with other reconstruction algorithms: the maximum likelihood (ML) EM with 100 iterations, maximum *a posteriori* (MAP)-EM with RDP with 100 iterations [12], and the hybrid DIP-based reconstruction proposed by Gong et al. with 200 iterations that performed two sub iterations of EM reconstruction and 10 sub-iterations of DIP update (same settings in the original implementation) [31].

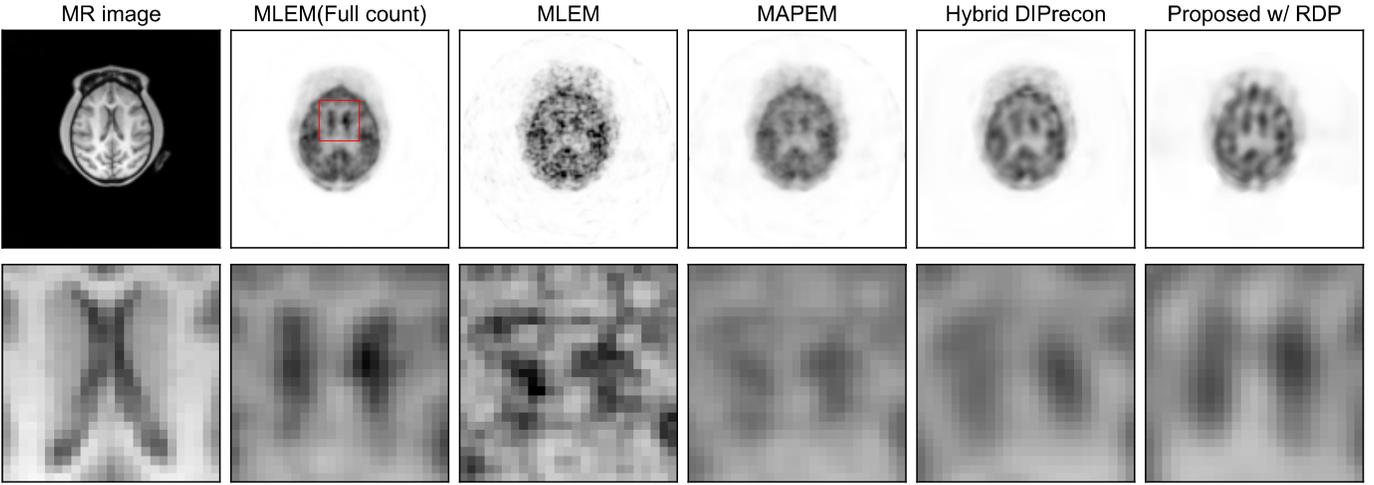

Fig. 9. Reconstruction results of the real preclinical brain [$^{18}$F]FDG data for different reconstruction algorithms; the MLEM, MAPEM [12], hybrid DIPrecon [34], and proposed method with RDP (Proposed w/ RDP) (left-to-right). The magnified images of the red squared region are shown in the bottom row.

The peak signal-to-noise ratio (PSNR) was calculated to evaluate reconstruction performance using the following equation:

$$\text{PSNR} = 10 \log_{10}\left(\frac{\max(K)^2}{\frac{1}{N}\|K - K'\|_2^2}\right), \quad (10)$$

where $N$ is the number of voxels, $K$ and $K'$ represent the ground truth and target reconstructed image, respectively, and $\max(\cdot)$ is the maximum value of the image.

The contrast recovery coefficient (CRC) and standard deviation (STD) were calculated using the following equations.

$$\text{CRC} = \left(\frac{\bar{a}}{\bar{b}} - 1\right) / \left(\frac{\bar{a}_{\text{gt}}}{\bar{b}_{\text{gt}}} - 1\right), \quad (11)$$

$$\text{STD} = \frac{1}{\bar{b}}\sqrt{\frac{1}{K_b}\sum_{k=1}^{K_b}(b_k - \bar{b})^2}, \quad (12)$$

where $\bar{a}_{gt}$ and $\bar{b}_{gt}$ are the ground truth uptakes in the inserted tumor and white matter regions, respectively. $\bar{a} = 1/K_a \sum_{k=1}^{K_a} a_k$ and $\bar{b} = 1/K_b \sum_{k=1}^{K_b} b_k$ are the mean uptakes of the inserted tumor in the $K_a$ regions and that of the white matter in the $K_b$ regions. $K_a$ and $K_b$ were set to 7 and 15, respectively.

## IV. RESULTS

### A. Simulation data

Fig. 2 and 3 show the reconstructed results of the simulation data with different regularization parameters $\beta$ and the effect of the regularization parameter for the proposed method with subsets 2. Finer brain structures and stable optimization results were obtained with $\beta = $ 5e-9; thus, this parameter value was used in this study. Fig. 4 shows the reconstruction results of the simulation data for different algorithms: MLEM, MAPEM with RDP, hybrid reconstruction, and the proposed method with and without RDP. The simulation results show that the proposed method with RDP improved PET image quality by reducing statistical noise and preserving the contrast of brain structures and inserted tumors. The highest PSNR of the proposed method with RDP supports the visual results.

Fig. 5 shows the curves of the CRC-STD tradeoff at the inserted tumors with different reconstruction algorithms. The proposed method with RDP achieved both higher quantitative accuracy in CRC and lower noise in STD than the other algorithms.

Fig. 6 shows the impact of RDP in terms of the PSNR for the proposed DIP-based reconstruction method. The proposed method with RDP improved the reconstruction performance more than the method without RDP. Furthermore, the proposed method with RDP did not reduce the quantitative performance, even with more iterations. Fig. 7 and 8 show the reconstructed results of the simulation data with different numbers of subsets and the effect of the number of subsets in terms of the PSNR for the proposed DIP-based reconstruction method. Note that a 3D implementation of end-to-end PET image reconstruction cannot work on current GPU boards unless the number of subsets is set to more than two owing to GPU memory limitations. The learning rate was adjusted empirically for each number of subsets. Similar trends were observed among these subsets, and the results indicate that the proposed block iteration algorithm was successfully demonstrated.

### B. Preclinical data

Fig. 9 shows the reconstruction results of preclinical real data for different algorithms with subsets 2. Compared with other reconstruction algorithms, our proposed reconstruction with RDP provides accurate putamen structures in low-dose PET imaging. Fig. 10 shows the line profiles through the putamen regions using different reconstruction algorithms. The best recovery of the putamen region was observed using the proposed method with RDP. These results indicate that the proposed method provides a more accurate quantitative solution for low-dose PET imaging.

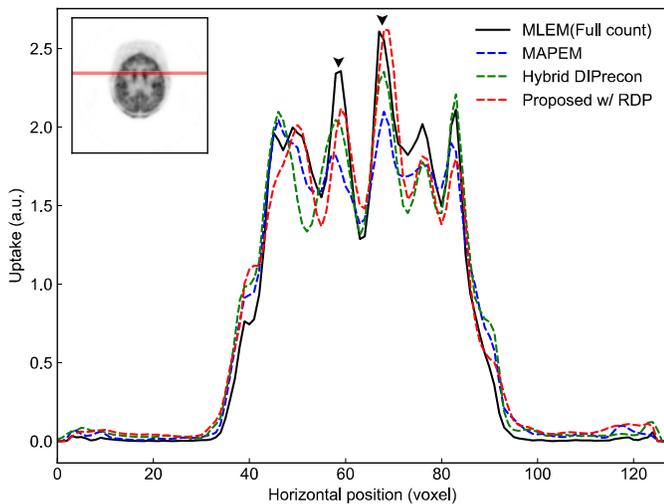

Fig. 10. Line profiles of the real preclinical brain [$^{18}$F]FDG data for different reconstruction algorithms; the MLEM, MAPEM [12], hybrid DIPrecon [34], proposed method, and proposed method with RDP (Proposed w/ RDP). The profile line is shown in the upper left. Triangular arrows are indicated the putamen regions.

## V. Discussion

In this study, we proposed an end-to-end DIP-based fully 3D PET image reconstruction incorporating a forward projection model into a loss function. To implement a practical end-to-end image reconstruction method for fully 3D PET, we modified the DIP optimization to block-iteration and sequentially learn an ordered sequence of block sinograms. In addition, we added a further penalty term for the RDP in the DIP optimization to enhance the quantitative accuracy of the reconstructed PET image and suppress overfitting problem.

According to the results in Fig. 2 and 3, the regularization parameter $\beta$ affects the smoothness of the reconstructed image. Thus, the RDP term can control the contrast and noise characteristics, similar to the MAPEM reconstruction algorithm. However, it is important to note that the scaling of $\beta$ is markedly different from that of the MAPEM because the MSE was used as the objective function in the proposed method (the negative log-likelihood was used for the MAPEM). For the simulation results, the proposed method without the RDP showed comparable results in terms of CRC-STD tradeoffs compared to the hybrid DIPrecon method. Hybrid DIPrecon uses the DIP and the conditional vector of the MR image under the same conditions as the proposed method - a similar tendency was observed in a previous report on 2D end-to-end DIP-based PET image reconstruction [36]. In contrast, the proposed method with RDP had the highest PET image quality by reducing statistical noise and preserving the contrast of brain structures and inserted tumors, compared to the hybrid DIPrecon method. Moreover, we found that the RDP term can boost the reconstructed image quality and achieve stable optimization because the graph in Fig. 6 shows a plateau at higher PSNR values. In other words, these results support that the RDP term suppresses the overfitting problem and eliminates the need for early stopping in DIP optimization; thus, it does not require strict monitoring of the iteration number.

For the preclinical experiment, the proposed method with RDP could recover the putamen uptake and structures despite the low-dose data of 1/20th of the full count data, whereas the other methods, including the hybrid DIPrecon method, could not recover it. These results indicate that the proposed method can be applied not only to simulation data but also to real data for low-dose PET imaging.

A block iterative algorithm is essential for the practical implementation of the proposed fully 3D PET reconstruction method. In general, the DIP framework does not perform mini-batch optimization, because it performs optimization on a single target image. In this method, to learn an ordered sequence of block sinograms, such as the ordered subset EM algorithm, we propose a block iterative algorithm as a mini-batch optimization. It is important to note that a 3D implementation of the end-to-end PET image reconstruction algorithm could not work on current GPU boards without using the proposed block iterative algorithm, owing to GPU memory limitations. In this method, GPU memory usage can be reduced by the number of subsets, for example, the size of the system matrix, which occupies most of the GPU memory usage in this study, can be reduced by 1/8 with subsets 8. The results in Fig. 7 show that the number of subsets did not markedly affect the tendency of the image quality in terms of PSNR, and our proposed method can be practically and easily used even on middle-end GPU boards.

The proposed end-to-end reconstruction method updates the trainable parameters of the neural network $\theta$ in one-step optimization using a deep learning framework, unlike the hybrid DIPrecon method, which separately optimizes two sub-problems, namely EM reconstruction and DIP denoising updates. Therefore, the proposed method does not require many hyperparameters. For example, in this method, the required number of hyperparameters is three, including the number of iterations, number of subsets, and learning rate. In contrast, the hybrid DIPrecon method requires five hyperparameters: the number of main iterations, two numbers of sub-iterations, the regularization parameter, and the learning rate. When using the RDP term, the reconstructed image quality can be easily adjusted using only the regularization parameter $\beta$, as shown in Fig. 2 and 3. These considerations indicate that the proposed method is simple and easy to implement for end-to-end, fully 3D PET image reconstruction.

The proposed method used the patient's MR image as an additional condition to enhance PET image quality, based on the success of previous reports [30,33,34,36]. A previous study showed poor noise characteristics when using random noise input for DIP-based PET image reconstruction, compared with MR image input [36]; thus, attention is needed to degrade noise characteristics when using random noise input in this study.

Most existing deep learning-based PET image reconstruction methods are data-driven approaches that incorporate trained networks that learn the relationship between large numbers of training pairs of high- and low-quality PET images. Compared to data-driven approaches, the proposed DIP-based reconstruction is free from the limitation of the quality of training datasets; that is to say, there are no performance restrictions with domain adaptation capabilities (e.g., different

PET probes, scanners, organs, and diseases) as well as an upper limit of high-quality images in the proposed method. Therefore, we believe that the proposed method, not only improves the performance of low-dose PET image reconstruction but also pushes the limits of the baseline for high-quality PET image reconstruction.

One limitation of this study is the learning stability when the number of subsets is increased. This may be due to the difficulties in optimizing the network from random initialization parameters. Using a pre-trained network [43,44] can easily solve the above-mentioned concern. In this study, we only evaluated the Monte Carlo simulation of the human [$^{18}$F]FDG brain and the preclinical dataset of the [$^{18}$F]FDG monkey brain. Further clinical evaluations, including different PET probes, PET scanners, organs, and diseases, will be included in our future work.

## VI. Conclusion

In this paper, we presented a first attempt to implement an end-to-end DIP-based fully 3D PET image reconstruction incorporating a forward projection model into a loss function. We evaluated the proposed reconstruction method using Monte Carlo simulation of human brain PET data and preclinical brain PET study with a rhesus monkey by comparing its results with those of the MLEM, MAPEM with RDP, and hybrid DIPrecon algorithms. The simulation study showed that the proposed method with RDP improves PET image quality by reducing statistical noise and preserving the contrast of brain structures and inserted tumors compared to other algorithms. In the preclinical study, finer structures and better contrast recovery were observed in the proposed method compared with the other reconstruction algorithms. These results indicate that the proposed method can produce high-quality images without a prior training dataset. Furthermore, the RDP term suppressed overfitting in the DIP optimization; thus, it frees the strict monitoring of the number of iterations. In conclusion, the proposed method is a key enabling technology for the straightforward and practical implementation of end-to-end DIP-based fully 3D PET image reconstruction.

## Acknowledgment

This work was supported by JSPS KAKENHI Grant Number JP22K07762.

## References

[1] M. E. Phelps, *PET: Molecular Imaging and Its Biological Applications*. New York, NY, USA: Springer, 2012.

[2] J. Dutta, R. M. Leahy, and Q. Li, "Non-local means denoising of dynamic PET images", *PLoS One*, vol. 8, no. 12, e81400, Dec. 2013.

[3] K. Ote *et al.*, "Kinetics-induced block matching and 5-D transform domain filtering for dynamic PET image denoising", *IEEE Trans. Radiat. Plasma Med. Sci.*, vol. 4, no. 6, pp. 720–728, Nov. 2020, doi: 10.1109/TRPMS.2020.3000221.

[4] B. T. Christian, N. T. Vandehey, J. M. Floberg et al., "Dynamic PET denoising with HYPR processing", *J. Nucl. Med.*, vol. 51, no. 7, pp. 1147–1154, 2010, doi: 10.2967/jnumed.109.073999.

[5] F. Hashimoto, H. Ohba, K. Ote, and H. Tsukada, "Denoising of dynamic sinogram by image guided filtering for positron emission tomography", *IEEE Trans. Radiat. Plasma Med. Sci.*, vol. 2, no. 6, pp. 541–548, Nov. 2018, doi: 10.1109/TRPMS.2018.2869936.

[6] A. J. Reader, and H. Zaidi, "Advances in PET image reconstruction", *PET Clin.*, vol. 2, no. 2, pp. 173-190, 2008, doi: 10.1016/j.cpet.2007.08.001.

[7] A. Alessio, et al., "Application and evaluation of a measured spatially variant system model of PET image reconstruction", *IEEE Trans. Med. Imaging*, vol. 29, no. 3, pp. 938-949, 2010, doi: 10.1109/TMI.2010.2040188.

[8] G. Wang, and J. Qi, "PET image reconstruction using kernel method", *IEEE Trans. Med. Imaging*, vol. 34, no. 1, pp. 61-71, Jan. 2015, doi: 10.1109/TMI.2014.2343916.

[9] T. Hebert, and R. Leahy, "A generalized EM algorithm for 3-D Bayesian reconstruction from Poisson data using Gibbs priors", *IEEE Trans. Med. Imaging*, vol. 8, no. 2, pp. 194-202, Jun. 1989, doi: 10.1109/42.24868.

[10] P. J. Green, "Bayesian reconstruction from emission tomography data using a modified EM algorithm", *IEEE Trans. Med. Imaging*, vol. 9, pp. 84-93, 1990, doi: 10.1109/42.52985.

[11] J. Qi, R. M. Leahy, S. R. Cherry, A. Chatziioannou, and T. H. Farquhar, "High-resolution 3D Bayesian image reconstruction using the microPET small-animal scanner", *Phys. Med. Biol.*, vol. 43, no. 4, pp. 1001, Apr. 1998, doi: 10.1088/0031-9155/43/4/027.

[12] J. Nuyts, D. Beque, P. Dupont, and L. Mortelmans, "A concave prior penalizing relative differences for maximum-a-posteriori reconstruction in emission tomography", *IEEE Trans. Nucl. Sci.*, vol. 49, no. 1, pp. 56-60, Feb. 2002, doi: 10.1109/TNS.2002.998681.

[13] X. Ouyang, W. H. Wong, V. E. Johnson, X. Hu, and C. T. Chen, "Incorporation of correlated structural images in PET image reconstruction", *IEEE Trans. Med. Imaging*, vol. 13, no. 4, pp. 627-640, Dec. 1994, doi: 10.1109/42.363105.

[14] B. Bai, Q. Li, and R. M. Leahy, "Magnetic resonance-guided positron emission tomography image reconstruction", *Seminars Nucl. Med.*, vol. 43, no. 1, pp. 30-44, Jan. 2013, doi: 10.1053/j.semnuclmed.2012.08.006.

[15] C. Comtat et al., "Clinically feasible reconstruction of 3D whole-body PET/CT data using blurred anatomical labels", *Phys. Med. Biol.*, vol. 47, no. 1, pp. 1, 2001, doi: 10.1088/0031-9155/47/1/301.

[16] A. J. Reader, et al., "Deep learning for PET image reconstruction", *IEEE Trans. Radiat. Plasma Med. Sci.*, vol. 5, no. 1, pp. 1–25, 2020, doi: 10.1109/TRPMS.2020.3014786.

[17] K. Gong, K. Kim, J. Cui, D. Wu, and Q. Li, "The Evolution of Image Reconstruction in PET: From Filtered Back-Projection to Artificial Intelligence", *PET Clin.*, vol. 16, no. 4, pp. 533–542, 2021, doi: 10.1016/j.cpet.2021.06.004.

[18] A. J. Reader, and G. Schramm, "Artificial intelligence for PET image reconstruction", *J. Nucl. Med.*, vol. 62, no. 10, pp. 1330-1333, 2021.

[19] J. Liu, et al., "Artificial Intelligence-Based Image Enhancement in PET Imaging: Noise Reduction and Resolution Enhancement", *PET Clin.*, vol. 16, no. 4, pp. 553–576, 2021, doi: 10.1016/j.cpet.2021.06.005.

[20] B. Zhu, J. Z. Liu, S. F. Cauley, B. R. Rosen, and M. S. Rosen, "Image reconstruction by domain-transform manifold learning", *Nature*, vol. 555, no. 7697, pp. 487–492, Mar. 2018, doi: 10.1038/nature25988.

[21] I. Häggström, C. R. Schmidtlein, G. Campanella, and T. J. Fuchs, "DeepPET: A deep encoder–decoder network for directly solving the PET image reconstruction inverse problem", *Med. Image Anal.*, vol. 54, pp. 253–262, May 2019, doi: 10.1016/j.media.2019.03.013.

[22] Z. Hu et al., "DPIR-Net: Direct PET image reconstruction based on the Wasserstein generative adversarial network", *IEEE Trans. Radiat. Plasma Med. Sci.*, vol. 5, no. 1, pp. 35–43, 2020, doi: 10.1109/TRPMS.2020.2995717.

[23] W. Whiteley, et al., "FastPET: near real-time reconstruction of PET histo-image data using a neural network", *IEEE Trans. Radiat. Plasma Med. Sci.* vol. 5, no. 1, pp. 65–77, 2020, doi: 10.1109/TRPMS.2020.3028364.

[24] T. Feng, et al., "Deep learning-based image reconstruction for TOF PET with DIRECT data partitioning format", *Phys. Med. Biol.*, vol. 66, no. 16, 165007, 2021, doi: 10.1088/1361-6560/ac13fe.

[25] K. Ote, and F. Hashimoto. "Deep-learning-based fast TOF-PET image reconstruction using direction information", Radiol. Phys. Technol., vol. 15, no. 1, pp. 72–82, 2022, doi: 10.1007/s12194-022-00652-8.

[26] K. Gong et al., "Iterative PET Image Reconstruction Using Convolutional Neural Network Representation", *IEEE Trans. Med. Imaging*, vol. 38, no. 3, pp. 675–685, Mar. 2019, doi: 10.1109/TMI.2018.2869871.

[27] A. Mehranian, and A. J. Reader, "Model-based deep learning PET image reconstruction using forward-backward splitting expectation maximization", *IEEE Trans. Radiat. Plasma Med. Sci.*, vol. 5, no. 1, pp. 54–64, 2020, doi: 10.1109/TRPMS.2020.3004408.

[28] D. Ulyanov, A. Vedaldi, and V. Lempitsky, "Deep image prior", *Int. J. Comput. Vis.*, vol. 128, no. 7, pp. 1867–1888, 2020, doi: 10.1007/s11263-020-01303-4.

[29] F. Hashimoto, H. Ohba, K. Ote, A. Teramoto, and H. Tsukada, "Dynamic PET Image Denoising Using Deep Convolutional Neural Networks Without Prior Training Datasets", *IEEE Access*, vol. 7, pp. 96594–96603, 2019, doi: 10.1109/ACCESS.2019.2929230.

[30] J. Cui et al., "PET image denoising using unsupervised deep learning", *Eur. J. Nucl. Med. Mol. Imaging*, vol. 46, no. 13, pp. 2780–2789, 2019, doi: 10.1007/s00259-019-04468-4.

[31] T. Yokota, K. Kawai, M. Sakata, Y. Kimura, and H. Hontani, "Dynamic PET Image Reconstruction Using Nonnegative Matrix Factorization Incorporated With Deep Image Prior", *Proc. IEEE Int. Conf. Comput. Vis.*, pp. 3126–3135, 2019, doi: 10.1109/ICCV.2019.00322.

[32] F. Hashimoto, et al., "4D deep image prior: dynamic PET image denoising using an unsupervised four-dimensional branch convolutional neural network", *Phys. Med. Biol.*, vol. 66, no. 1, Jan. 2021, doi: 10.1088/1361-6560/abcd1a.

[33] Y. Onishi, et al., "Anatomical-guided attention enhances unsupervised PET image denoising performance", Med. Image Anal., vol.74, 102226, 2021, doi: 10.1016/j.media.2021.102226.

[34] K. Gong, C. Catana, J. Qi, and Q. Li, "PET image reconstruction using deep image prior", *IEEE Trans. Med. Imaging*, vol. 38, no. 7, pp. 1655–1665, July 2019, doi: 10.1109/TMI.2018.2888491.

[35] K. Ote, F. Hashimoto, Y. Onishi, T. Isobe, and Y. Ouchi, "List-Mode PET Image Reconstruction Using Deep Image Prior", *arXiv preprint*, arXiv:2204.13404, 2022, doi: 10.48550/arXiv.2204.13404.

[36] F. Hashimoto, K. Ote, and Y. Onishi, "PET Image Reconstruction Incorporating Deep Image Prior and a Forward Projection Model", *IEEE Trans. Radiat. Plasma. Med. Sci.*, vol. 6, no. 8, pp. 841–846, Nov. 2022, doi: 10.1109/TRPMS.2022.3161569.

[37] E.J. Teoh, D.R. McGowan, R.E. Macpherson, K.M. Bradley, and F.V. Gleeson, "Phantom and Clinical Evaluation of the Bayesian Penalized Likelihood Reconstruction Algorithm Q.Clear on an LYSO PET/CT System", J. Nucl. Med., vol. 56, no. 9, pp. 1447–1452, 2015, doi: 10.2967/jnumed.115.159301.

[38] Ö. Çiçek, A. Abdulkadir, S. S. Lienkamp, T. Brox, and O. Ronneberger, "3D U-Net: Learning dense volumetric segmentation from sparse annotation", *Proc. Int. Conf. Med. Image Comput. Comput.-Assist. Intervent*, pp. 424-432, Oct. 2016, doi: 10.1007/978-3-319-46723-8_49.

[39] C. Zhu, R. H. Byrd, P. Lu, and J. Nocedal, "Algorithm 778: L-BFGS-B: Fortran subroutines for large-scale bound-constrained optimization", *ACM Trans. Math. Softw.*, vol. 23, no. 4, pp. 550–560, Dec. 1997, doi: 10.1145/279232.279236.

[40] S. Yatawatta, L. De Clercq, H. Spreeuw, and F. Diblen, "A stochastic LBFGS algorithm for radio interferometric calibration", *Proc. IEEE Data Science Workshop (DSW)*, pp. 208–212, June 2019, doi: 10.1109/DSW.2019.8755567.

[41] Y. Onishi, et al., "Performance evaluation of dedicated brain PET scanner with motion correction system", Ann. Nucl. Med., vol. 36, no. 8, pp. 746–755, 2022, doi: 10.1007/s12149-022-01757-1.

[42] M. Lubberink, T. Kosugi, H. Schneider, H. Ohba, and M. Bergström, "Non-stationary convolution subtraction scatter correction with a dual-exponential scatter kernel for the Hamamatsu SHR-7700 animal PET scanner", vol. 49, no. 5, pp. 833–842, 2004, doi: 10.1088/0031-9155/49/5/013.

[43] J. Cui, et al., "Populational and individual information based PET image denoising using conditional unsupervised learning", *Phys. Med. Biol.*, vol. 66, no. 15, 155001, 2021, doi: 10.1088/1361-6560/ac108e.

[44] Y. Onishi, F. Hashimoto, K. Ote, K. Matsubara, and M. Ibaraki, "Using Self-Supervised Pretraining Model for Unsupervised PET Image Denoising", *IEEE Nucl. Sci. Symp. Med. Imag. (NSS/MIC)*, Milano, Italy, 2022.